# Attention Guided Conditional Adversarial Learning for EMG Artifact Suppression in Single Channel EEG

Haoyi Wang[1], Haowei Wang[2], Yihang Li[2], Wenjie Zhang[3], and Yibo Wang[1]

[1] Department of Electrical Engineering, Columbia University, New York, NY 10027, USA

[2] School of Communication and Information Engineering, Xi'an University of Posts and Telecommunications, Xi'an 710121, China

[3] Faculty of Agriculture, The University of Tokyo, Tokyo 113-8657, Japan

***Abstract-****Electromyographic activity can dominate scalp electroencephalography across a broad frequency range and obscure neural structure that is relevant to analysis and brain computer interfaces. This paper evaluates P2P Trans, a conditional adversarial denoiser that combines a U shaped encoder decoder, multi head self attention at the bottleneck, cross attention guided skip fusion, and a PatchGAN discriminator. The reported experiment used EEGdenoiseNet segments represented as 1024 sample sequences, synthesized paired mixtures at ten signal to noise ratios from minus 7 to 2 dB, and evaluated 50000 training examples and 5980 test examples. Against five convolutional encoder decoder baselines, P2P Trans achieved the best mean correlation coefficient of 0.8645 and the lowest mean spectral RRMSE of 0.3596. Its mean temporal RRMSE was 0.5797 and did not lead the comparison, which indicates a tradeoff between spectral suppression and pointwise waveform fidelity. The advantage was strongest under severe contamination. The results support attention guided adversarial restoration as a useful low SNR approach, while also identifying the need for subject independent evaluation, direct one dimensional modeling, and downstream task validation.*

## I INTRODUCTION

Scalp electroencephalography (EEG) offers millisecond scale temporal resolution, but its small amplitude makes it sensitive to physiological interference. Electromyographic (EMG) artifacts are especially difficult because muscle activity is nonstationary, varies across recording sites, and overlaps the beta and gamma bands used in many neural analyses [1], [2]. Residual EMG can therefore distort spectral estimates and reduce the reliability of downstream classification.

Regression, adaptive filtering, independent component analysis, and canonical correlation analysis remain useful when reference channels or statistical separation assumptions are available [2], [11], [12]. Single channel and low density systems often lack that information. End to end networks instead learn a mapping from contaminated segments to clean targets, as demonstrated by EEGdenoiseNet and later convolutional and Transformer based methods [3]-[7]. Their main challenge is to suppress broadband muscle energy without smoothing transient neural activity.

P2P Trans addresses this problem with a conditional adversarial objective. A U shaped generator extracts local patterns, self attention models long range relationships at the compressed representation, and cross attention filters skip features before reconstruction. A patch discriminator penalizes locally implausible outputs. The study makes three contributions: it evaluates the hybrid architecture across ten contamination levels, reports both time and frequency domain error, and analyzes the observed tradeoff rather than reducing performance to a single score.

## II RELATED WORK

### *A Classical and neural artifact suppression*

Classical component separation can work well with sufficient channels, but automated component identification and assumption violations remain practical limitations [11], [12], [14]. One dimensional residual CNNs avoid manual features and preserve the native sequence structure [4]. Segmentation denoising networks further separate artifact localization from waveform restoration [5]. These models are efficient, although a finite convolutional receptive field can make global context difficult to represent.

### *B Attention and adversarial restoration*

Pix2pix established conditional adversarial learning for paired translation [8], while the Transformer introduced content dependent interactions over long ranges [9]. Hybrid U Net and attention designs combine local inductive bias with global context [10], and adversarial modeling has also been used to augment BCI data [13]. Recent EEG denoisers use Transformer blocks or parallel CNN and Transformer branches to model complementary dependencies [6], [7], [15]. P2P Trans differs by placing self attention at the bottleneck and using cross attention to filter encoder features before they reach the decoder.

More recent work has emphasized frequency aware routing and compact architectures [16], [17]. Those developments motivate the future directions in Section VI, but they were not part of the experiment reported here and are not used to reinterpret its numerical results.

## III DATA AND PROBLEM FORMULATION

### *A Dataset construction*

The experiment used the public EEGdenoiseNet benchmark [3]. The reported preprocessing pipeline represented each two second segment with 1024 samples. It contained 4514 clean EEG segments and 5598 EMG segments. Clean EEG was resampled with replacement to match the EMG count, after which 5000 base pairs were assigned to training and 598 to testing. Each pair was mixed at every integer SNR from minus 7 through 2 dB, producing 50000 training examples and 5980 test examples.

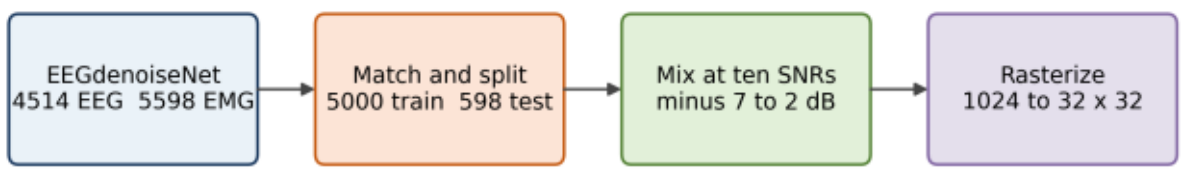

Fig. 1. Construction of the paired benchmark used in the reported experiment.

For clean EEG x and an EMG segment n, the contaminated signal y was formed with an amplitude coefficient selected for the desired SNR. Equation 1 states the mixing rule using the conventional power ratio definition.

$$y = x + \lambda n, \qquad \lambda = \frac{\mathrm{RMS}(x)}{\mathrm{RMS}(n)} 10^{-\mathrm{SNR}/20} \tag{1}$$

*B Rasterized signal representation*

Each 1024 sample sequence was scaled and reshaped into a 32 by 32 grayscale array. This operation is a rasterization of the time series, not a short time Fourier transform. It preserves all sample values but introduces two dimensional neighborhoods that do not all correspond to adjacent time points. The generator output was mapped back to the clean signal amplitude range and flattened to one dimension for evaluation.

TABLE I
EXPERIMENTAL DATA CONFIGURATION

| Item | Value |
|---|---|
| Clean EEG | 4514 segments |
| EMG | 5598 segments |
| Representation | 1024 samples 32 x 32 |
| SNR | minus 7 to 2 dB |
| Training | 50000 mixtures |
| Testing | 5980 mixtures |

## IV P2P TRANS ARCHITECTURE

*A Attention guided generator*

The generator follows an encoder decoder topology with channels increasing from 64 to 512. Convolution and batch normalization blocks form the encoder. Multi head self attention at the bottleneck connects positions throughout the compressed feature map. For query Q, key K, and value V, attention is computed as follows.

$$\mathrm{Attention}(Q, K, V) = \mathrm{softmax}\left(\frac{QK^{\mathrm{T}}}{\sqrt{d_k}}\right)V \tag{2}$$

Decoder features provide the queries for multi head cross attention, while encoder skip features provide keys and values. The resulting weights gate shallow features before concatenation. This design aims to retain waveform details that agree with global context and suppress skip features dominated by muscle activity. Upsampling blocks reconstruct a single channel 32 by 32 output followed by a hyperbolic tangent activation.

*B Conditional discriminator and objective*

The discriminator receives the contaminated map paired with either the clean target or the generated output. Four two dimensional convolution blocks classify overlapping local patches. Adversarial supervision encourages realistic local structure, and an L1 term anchors the prediction to its paired target.

$$\mathcal{L}_G = \mathcal{L}_{\mathrm{cGAN}}(G, D) + \lambda_1 \lVert x - G(y) \rVert_1 \tag{3}$$

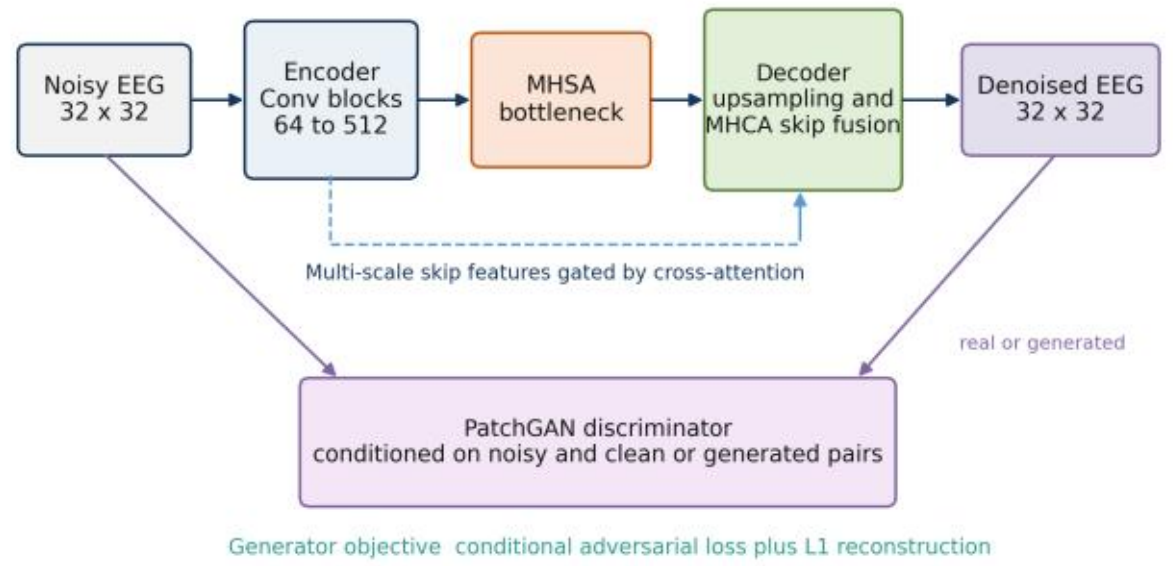

Fig. 2. P2P Trans generator and conditional PatchGAN training path.

## V EVALUATION AND RESULTS

*A Metrics and baselines*

Evaluation used the correlation coefficient CC, temporal relative root mean square error RRMSE t, and spectral relative root mean square error RRMSE f. With clean signal x and estimate x hat, the time domain metrics are defined in Equation 4.

$$\mathrm{RRMSE}_t = \frac{\mathrm{RMS}(\hat{x} - x)}{\mathrm{RMS}(x)}, \qquad \mathrm{CC} = \frac{\mathrm{Cov}(\hat{x}, x)}{\sqrt{\mathrm{Var}(\hat{x})\mathrm{Var}(x)}} \tag{4}$$

Spectral error compares power spectral density estimates over the analyzed band.

$$\mathrm{RRMSE}_f = \frac{\mathrm{RMS}(\mathrm{PSD}(\hat{x}) - \mathrm{PSD}(x))}{\mathrm{RMS}(\mathrm{PSD}(x))} \tag{5}$$

P2P Trans was compared with FPN, U Net, MCGUNet, LinkNet, and MultiResUNet3 under the same ten SNR conditions. Higher CC is better, whereas lower RRMSE indicates a closer reconstruction.

*B Aggregate comparison*

Table II reports the average over all SNR levels. P2P Trans produced the highest mean CC and reduced spectral RRMSE by 28.2 percent relative to the next lowest value, 0.5011 from MCGUNet. It did not minimize temporal RRMSE: its value of 0.5797 was 4.0 percent above the MCGUNet result. The comparison therefore supports a spectral and low SNR advantage, not uniform dominance across every metric.

TABLE II
AVERAGE DENOISING PERFORMANCE ACROSS SNR LEVELS

| Model | CC up | RRMSE t down | RRMSE f down |
|---|---|---|---|
| FPN | 0.7995 | 0.5630 | 0.5234 |
| U Net | 0.7927 | 0.5647 | 0.5232 |
| MCGUNet | 0.8023 | 0.5573 | 0.5011 |
| LinkNet | 0.7924 | 0.5662 | 0.5239 |
| MultiResUNet3 | 0.8028 | 0.5594 | 0.5056 |
| **P2P Trans** | **0.8645** | **0.5797** | **0.3596** |

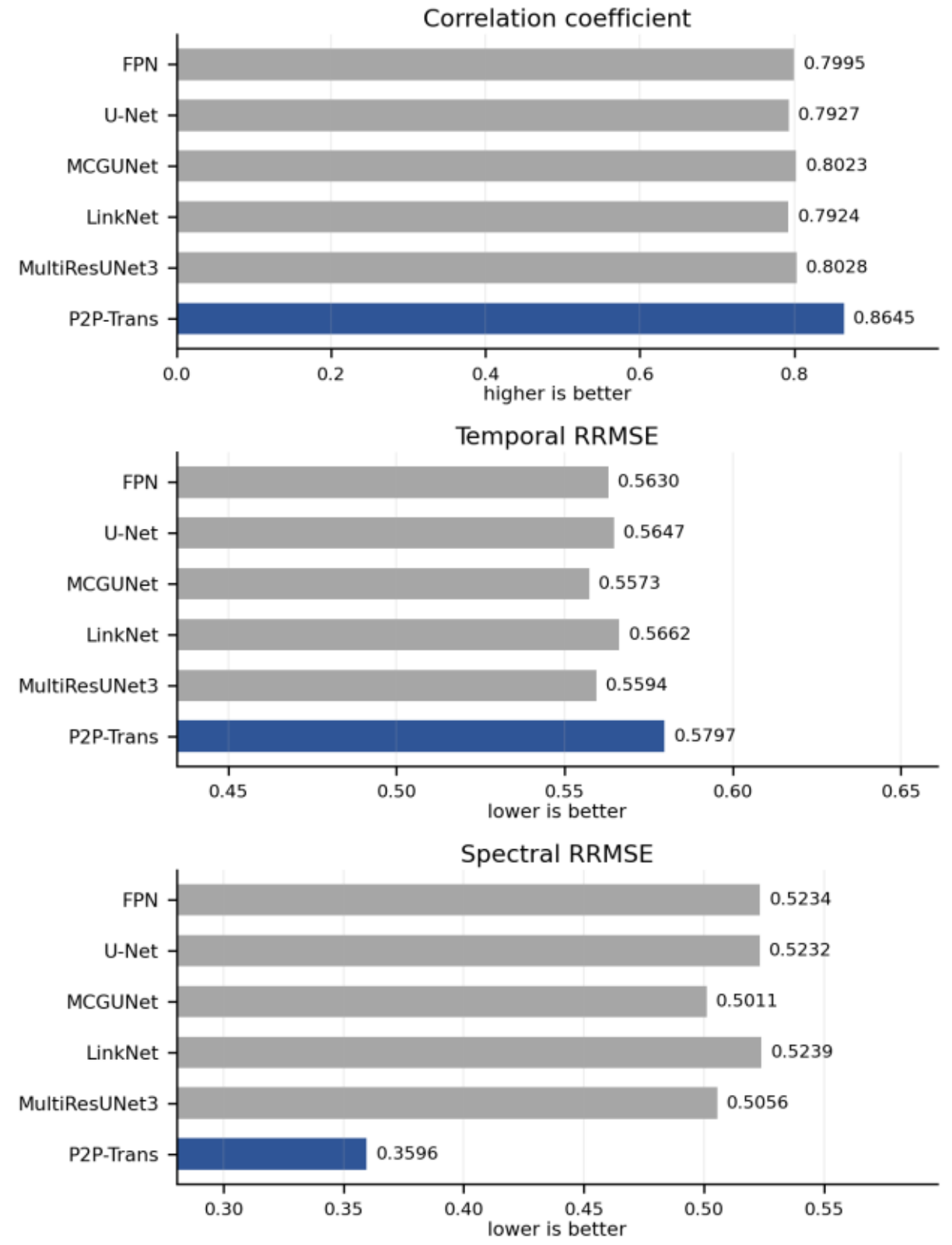

Fig. 3. Mean performance reported in Table II. Blue bars denote P2P Trans.

### C Behavior across contamination levels

The reported SNR sweep shows that the correlation advantage was concentrated in the hardest mixtures. At minus 7 dB, P2P Trans reached a CC of approximately 0.8247 while the convolutional baselines were below 0.60. As contamination decreased, the baseline CC values rose more quickly and overtook P2P Trans near the high SNR end. Spectral RRMSE remained lower for the proposed model across the sweep and reached approximately 0.2909 at 2 dB. These trends are consistent with aggressive suppression of broadband EMG energy, accompanied by some loss of pointwise detail when the input is already relatively clean.

## VI DISCUSSION

### A Interpretation

Self attention gives the bottleneck access to nonlocal structure, and cross attention prevents every encoder feature from passing unchanged through the skip connections. Patch based adversarial feedback adds a local realism constraint. Together, these mechanisms provide a plausible explanation for the strong low SNR correlation and spectral error. The weaker temporal RRMSE shows that frequency recovery and samplewise fidelity are related but distinct objectives.

### B Limitations

The benchmark uses synthetic additive mixtures and does not reproduce every interaction found in naturally recorded muscle contaminated EEG. The reported experiment also does not document a subject independent partition, optimizer settings, random seeds, or a complete ablation of self attention, cross attention, and adversarial loss. The 32 by 32 rasterization creates artificial spatial adjacency. Finally, reconstruction metrics do not establish that denoising improves a downstream BCI or clinical task [17]. These constraints limit the strength and reproducibility of the claim.

### C Current technical directions

A follow up study should operate directly on one dimensional signals or combine a temporal branch with a true time frequency branch. Multi resolution spectral loss could explicitly preserve delta through gamma structure, while linear attention or compact Conformer blocks could reduce computational cost. Frequency aware routing [16], self supervised pretraining, uncertainty estimates, and subject adaptive fine tuning are also relevant. Each change requires a new controlled experiment; none is included in the numerical results above.

## VII CONCLUSION

P2P Trans combines conditional adversarial learning with self attention and attention filtered skip fusion for EMG artifact suppression. On the reported EEGdenoiseNet mixtures, it achieved the best mean CC and spectral RRMSE among six compared models, with its clearest advantage under severe contamination. The higher temporal RRMSE and incomplete reproducibility record prevent a claim of universal superiority. Future validation should use subject independent splits, natural artifacts, direct temporal modeling, ablation studies, and downstream neural decoding outcomes.

## AUTHOR CONTRIBUTIONS

Haowei Wang led the current project coordination and preparation of this manuscript. Haoyi Wang developed and evaluated the original P2P Trans system and contributed the underlying experimental analysis.